\documentclass{PoS}

\newcommand{\beqn}{\begin{eqnarray}}
\newcommand{\eeqn}{\end{eqnarray}}
\newcommand{\be}{\begin{equation}}
\newcommand{\ee}{\end{equation}}

\newcommand{\ba}{\begin{array}{c}}
\newcommand{\bat}{\begin{array}{cc}}
\newcommand{\ea}{\end{array}}

\newcommand{\bi}{\begin{itemize}}
\newcommand{\ei}{\end{itemize}}

\newcommand{\rcht}{R$\chi$T}

\newcommand{\Frac}[2]{\frac{\displaystyle #1}{\displaystyle #2}}

\newcommand{\cO}{{\cal O}}

\newcommand{\mL}{\mathcal{L}}

\newcommand{\lsim}{\stackrel{<}{_\sim}}
\newcommand{\gsim}{\stackrel{>}{_\sim}}

\ShortTitle{Weinberg sum rules at NLO in $1/N_C$}

\abstract{
We study the relevance of different renormalization schemes
in Resonance Chiral Theory. The  $SS-PP$ correlator is explicitly computed
at the one-loop level.
Demanding the operator product expansion behaviour at short distances
produces a new set of constraints, as some logarithmic terms are absent at
high energies. Likewise, the loops induce subleading corrections in $1/N_C$
to the leading-order constraints, the Weinberg sum rules.
We find that the short-distance conditions from a minimally subtracted scheme
generate large uncertainties which, alternatively, can be largely
simplified in other schemes.}
\FullConference{
Sixth International Workshop on Chiral Dynamics\\ July 6-10 2009,
Bern, Switzerland.
\\
\vspace*{0.5cm} This work was supported in part by the Center for
Particle Physics (Project no. LC 527), GAUK (Project no.6908;
114-10/258002), CICYT-FEDER-FPA2008-01430, SGR2005-00916, the
Spanish Consolider-Ingenio 2010 Program CPAN (CSD2007-00042), the
Juan de la Cierva Program  and the EU Contract No.
MRTN-CT-2006-035482, \lq\lq FLAVIAnet''. J.~T. is supported by the
U.S.Department of State (International Fulbright S\&T award). }

\title{Weinberg sum rules at NLO in $1/N_C$}
\author{Juan Jos\'e Sanz-Cillero$^{\,a}$ and
\speaker{Jaroslav Trnka}$^{\,bc}$ \\
$^a$ Grup de F\'\i sica Te\`orica and IFAE, Univ.
Aut\`onoma de Barcelona, 08193 Barcelona, Spain\\
$^b$ Charles University,
Institute of Particle and Nuclear Physics, V Hole\v{s}ovi\v{c}k\'ach 2,
18000 Prague, Czech Republic\\
$^c$ Department of Physics, Princeton University, 08540 Princeton, New
Jersey, USA\\
E-mail: jtrnka@princeton.edu}

\renewcommand{\logo}
{\parbox{\textwidth}{\begin{flushright} \vspace*{2cm}
 PUPT-2328
\end{flushright}\vspace*{-0.3cm}
\includegraphics{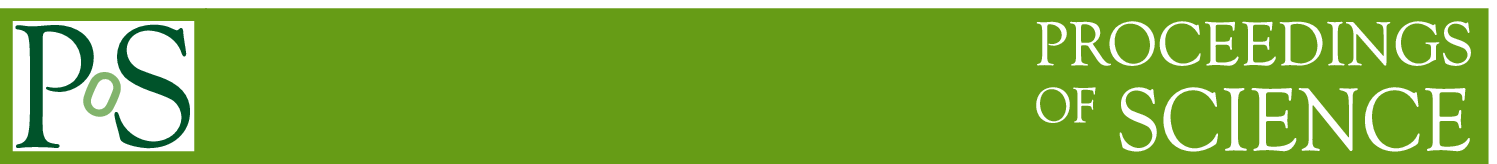}}
}

\begin{document}

\section{Introduction}

The effective field theory (EFT) approach is a very powerful tool
for the investigation of Quantum Chromodynamics (QCD) at long distances.
Chiral Perturbation theory ($\chi$PT)~\cite{Weinberg1,Gasser1} is the EFT
for the description of the chiral (pseudo) Goldstones in the low energy domain
$E\ll \Lambda_H\sim 1\,\rm{GeV}$, with $\Lambda_H$ typically the scale of
the lowest resonance masses. Recent progress has allowed to carry $\chi$PT
up to ${\cal O}(p^6)$, i.e., up to the two-loop level~\cite{Bijnens}.

In the intermediate resonance region, $\Lambda_H \lsim E \lsim  2 GeV$,
$\chi$PT stops being valid and one must explicitly include
the resonance fields  in the Lagrangian description.
Resonance Chiral Theory (R$\chi$T)  describes the interaction
of resonance and pseudo-Goldstones within a general chiral invariant
framework~\cite{Ecker1, Ecker2}. Alternatively to the chiral counting,
it uses the $1/N_C$ expansion of QCD in the limit of large number of colours~\cite{NC}
as a guideline to organize the perturbative expansion.
At leading order (LO), just tree-level diagrams contribute
while loop diagrams yield higher order effects.

The infinite tower of mesons contained in large--$N_C$ QCD is often truncated
to the lowest states in each channel, the so called single
resonance approximation (SRA).
This approximation has led to successful predictions
of ${\cal O}(p^4)$ and  ${\cal O}(p^6)$ low-energy constants
(LECs)~\cite{Ecker1,Ecker2,Pich,Karol,Estim}.
However, the study of Regge models with an infinite number
of mesons has shown that if one keeps just the lightest states
with exactly the same couplings and masses of the full model then
one get wrong values for the LECs~\cite{L8-Peris}.
Likewise, that analysis finds that the truncated theory do not produce the right
short-distance (SD)  behaviour.
Thus, in a matching with the OPE power behaviour the parameters
of the truncated theory will become shifted in order to accommodate
the right short-distance dependence.
Chiral symmetry ensures the proper low-momentum structure of the \rcht\ amplitudes
around $p^2=0$  but their high energy behaviour is not fixed by symmetry alone.
Nevertheless, one knows that for large Euclidean momenta, $(-p^2)\gsim 2$~GeV$^2$
the $SS-PP$ correlator is expected to follow a vanishing behaviour prescribed
by  the OPE.   In that sense, the matched amplitude can be understood
with the help of Padé
approximants as a rational interpolator between the deep Euclidean $p^2=-\infty$
and the low-energy domain around $p^2=0$~\cite{MHA,Pade}.
The Weinberg sum-rules
(WSR)~\cite{Weinberg2} yield the most convenient parameters for the interpolation
rather than accurate determinations of the resonance couplings.

Not much is known about the extension of R$\chi$T beyond the tree
level approximation. Although some theoretical issues on the
renormalizability of \rcht\ still need further
clarification~\cite{prop},   several chiral LECs have been already
computed up to NLO in $1/N_C$ through QFT one-loop
calculations~\cite{QFT}, dispersion relations~\cite{form} and even
analyzed with the help of renormalization group
techniques~\cite{Group}. Here we present the basic ideas of the work
in Ref.~\cite{Our}, where the $SS-PP$ correlator is computed up to
next-to-leading order in $1/N_C$ (NLO).  The one-loop amplitude is
then taken as an improved interpolator between long and short
distances and the corresponding  modifications to the former WSRs
are extracted. The amplitude is first computed within the
subtraction scheme of $\chi$PT~\cite{Gasser1}.  However, though
equivalent at low energies, some appropriate schemes are found to be
more convenient and to introduce less uncertainties in the SD
constraints.

\section{Weinberg sum rules at leading and next-to-leading order}

The two-point Green function $SS-PP$ we are interested in is defined by
\begin{equation}
\Pi_{S-P}^{ab}(p) \,\,=\,\, i\,\int d^4x e^{ip\cdot x}
\langle 0| T[S^a(x)S^b(0)-P^a(x)P^b(0)]|0\rangle
= \delta^{ab}\Pi(p^2)\,,
\end{equation}
with $S^a=\bar{q}\frac{\lambda_a}{\sqrt{2}} q$ and
$P^a=i\bar{q}\frac{\lambda_a}{\sqrt{2}}\gamma_5 q$,
being $\lambda_a$ the Gellmann matrices ($a=1,\dots 8$).

For convenience, the  R$\chi$T Lagrangian can be organized in the form
${\cal L} = {\cal L}_{GB} + {\cal L}_R + {\cal L}_{RR'} + \dots \, $,
where ${\cal L}_{GB}$ contains just Goldstone bosons and external sources,
${\cal L}_{R}$ includes operators with also one resonance field $R$, etc.
$\mL_{GB}$ is provided by the $\cO(p^2)$
$\chi$PT operators and the terms with one resonce field are given in the
SRA by~\cite{Ecker1}
\begin{equation}
{\cal L}_{R} = \frac{F_V}{2\sqrt2}\langle V_{\mu\nu}
f_+^{\mu\nu}\rangle + \frac{iG_V}{2\sqrt2}\langle V_{\mu\nu}
[u^\mu,u^\nu]\rangle +\frac{F_A}{2\sqrt2}\langle A_{\mu\nu}
f_-^{\mu\nu}\rangle + c_d\langle S u^\mu u_\mu \rangle + c_m \langle S
\chi_+\rangle + id_m \langle P\chi_-\rangle,
\label{Lagr2}
\end{equation}
where at tree-level operators with two or more resonances do not contribute.

If one  computes the one-loop correlator, the perturbative result shows the
form~\cite{form}
\begin{equation}
\Frac{1}{B_0^2}\, \Pi (p^2) \, =\,  \frac{2F^2}{p^2} \, +\,
\frac{16 c_{m}^2}{M_{S}^2-p^2}\, -\, \frac{16 d_{m}^2}{ M_{P}^2-p^2}
\,\,\,+\,\,\, \rho(p^2)
\, ,
\label{LO}
\end{equation}
with $\rho(p^2)$ containing the renormalized loop contributions and
other tree-level contributions subleading in $1/N_C$~\cite{form,Our}.
The correlator has then the high-energy expansion~\cite{NLOsatura},
\begin{equation}
\frac{1}{B_0^2}\Pi(p^2) \,\,\,=\,\,\,  \sum_{k=0,2,4...}
\Frac{1}{p^k}\,\, \left( \alpha_k^{(p)} +
\alpha_k^{(\ell)}\ln \frac{-p^2}{\mu^2}\right) \, .
\end{equation}
The requirement that the amplitude follows the high energy OPE behavior~\footnote{The
tiny dimension four condensate
$\frac{1}{B_0^2}\, \langle \cO_{(4)}^{^{SS-PP}}\rangle\simeq
- 12 \pi\alpha_S F^4$ will be neglected in this work~\cite{Pich,Peris}.
}
$\Pi(p^2)\stackrel{p^2\rightarrow\infty}{\longrightarrow} 1/p^6$ produces
the SD constraints~\cite{NLOsatura} for the log terms
$\alpha_0^{(\ell)}=\alpha_2^{(\ell)}=\alpha_0^{(\ell)}=0$, and the non-logarithmic
conditions $\alpha_0^{(p)}=0$ and
\begin{eqnarray}
\alpha_2^{(p)}  &=& 2F^2+ 16 d_m^2-16c_m^2
+ A(\mu)=0,
\qquad\,\,\, \alpha_4^{(p)} =
16 d_m^2 M_P^{2}  -16 c_m^2 M_S^2
+ B(\mu)=0\, .
\label{WSR2}
\end{eqnarray}
At LO in $1/N_C$ there are no logs ($\alpha^{(\ell)}_k=0$). The
remaining non-logarithmic constraints require the absence of local
terms ($\alpha_0^{(p)}=0$) and the  usual (large--$N_C$)  Weinberg
sum-rules ${  8\, c_m^2-8\, d_m^2-F^2=0  }$, ${
c_m^2M_S^2-d_m^2M_P^2=0  }$~\cite{Pich,Weinberg2}

At NLO, the WSRs gain the subleading corrections $A(\mu)$ and
$B(\mu)$~\cite{form,Our}~\footnote{ If one considers just the \rcht\
Lagrangian $\mL_{GB}+\mL_{R}$~\cite{Ecker1}, the NLO terms $A(\mu)$
and $B(\mu)$ result~\cite{Our}
\begin{eqnarray}
A(\mu)  &=&- \frac{3d_m^2M_P^2}{\pi^2F^2}\left(\ln\frac{M_P^2}{\mu^2}-1\right)
+\frac{3c_m^2M_S^2}{\pi^2F^2}\left(\ln\frac{M_S^2}{\mu^2}-1\right)
+\frac{6c_d^2c_m^2M_S^2}{\pi^2F^4}
\nonumber\\
&& \qquad
-\frac{6c_dc_mM_S^2}{\pi^2F^2}\left(\ln\frac{M_S^2}{\mu^2}+\frac14\right)
+\frac{9c_d^2M_S^2}{4\pi^2F^2}\left(\ln\frac{M_S^2}{\mu^2}+\frac12\right)
+\frac{9G_V^2M_V^2}{8\pi^2F^2}\left(\ln\frac{M_V^2}{\mu^2}+\frac12\right)  \,,
\nonumber\\
B(\mu)&=&
-\frac{3d_m^2M_P^4}{2F^2\pi^2} +\frac{9c_d^2c_m^2M_S^4}{F^4\pi^2}
+\frac{3c_m^2M_S^4}{2F^2\pi^2}-\frac{6c_dc_mM_S^4}{\pi^2F^2}
-\frac{9c_d^2M_S^4}{4\pi^2F^2}\left(\ln\frac{M_S^2}{\mu^2}-\frac12\right)
\nonumber\\
&&\qquad
+\frac{3c_dc_mM_S^4}{\pi^2F^2}\ln\frac{M_S^2}{\mu^2}
-\frac{9G_V^2M_V^4}{8\pi^2F^2}\left(\ln\frac{M_V^2}{\mu^2}- \frac12\right) \, .
\nonumber
\end{eqnarray}
}.
Notice that now the couplings
in~(\ref{WSR2})  are the renormalized ones.

One can then consider  a different renormalization scheme for
$\kappa=c_m, \, d_m,\, M_S,\, M_P$  (denoted with hat in the new
scheme). The difference between the two schemes would be provided by
the shifts
$  \kappa = \hat{\kappa} + \Delta \kappa  $,
with  $\Delta\kappa$ a finite constant formally subleading.
Since  $A(\mu)$ and $B(\mu)$ are already NLO in~(\ref{WSR2}),
their variation  is sub-subdominant and can be neglected, leaving
\begin{eqnarray}
&&\alpha_2^{(p)}  = 2F^2+ 16 \, \hat{d}_m^2-16 \, \hat{c}_m^2 \,\,\,
+\,\,\,\left[ 32  \, \hat{d}_m \Delta d_m -32  \, \hat{c}_m \Delta c_m\,\,+\,\,  A(\mu)
\right]\,\,=\,\ 0 \,,  \label{eq.WSR+scheme}
\\
&&\alpha_4^{(p)} =
16   \hat{d}_m^2 \hat{M}_P^{2}  -16   \hat{c}_m^2 \hat{M}_S^2
+ \left[  32   \hat{M}_P^2 \hat{d}_m \Delta d_m  + 16   \hat{d}_m^2 \Delta M_P^2
-32    \hat{M}_S^2 \hat{c}_m \Delta c_m
-  16    \hat{c}_m^2 \Delta M_S^2 +  B(\mu)
\right]  = 0  .
\nonumber
\end{eqnarray}
The terms within the brackets, $[ \cdots ]$, correspond to the
finite renormalized contributions from the one-loop diagrams in the
new scheme. In general, one finds that the expressions in the
brackets suffer from large numerical uncertainties, depending on the
precise values of the resonance couplings. However, there is a
convenient scheme where the expressions in the brackets become zero.
In that case, (\ref{eq.WSR+scheme}) shows the same structure of the
large--$N_C$ WSRs~\cite{Pich}, though now in terms of renormalized
parameters $\hat{\kappa}$.   Furthermore, the change of scheme does
not change the low-energy prediction for the LECs~\cite{Our}. It
just  removes the former uncertainty in the NLO  high-energy
constraints~(\ref{WSR2}).



\begin{thebibliography}{99}
\bibitem{Weinberg1} S.~Weinberg, Physica A \textbf{96} (1979) 327.

\bibitem{Gasser1} J.~Gasser and H.~Leutwyler, Annals Phys.\ \textbf{158}
(1984) 142; Nucl.\ Phys.\ B \textbf{250} (1985) 465.

\bibitem{Bijnens} J.~Bijnens, Prog.\ Part.\ Nucl.\ Phys.\ \textbf{58} (2007) 521;
[arXiv:0904.3713 [hep-ph]];  talk at this conference.

\bibitem{Ecker1}
    G.~Ecker, J.~Gasser, A.~Pich and E.~de Rafael,
    Nucl.\ Phys.\  B {\bf 321} (1989) 311.

\bibitem{Ecker2}
    G.~Ecker, J.~Gasser, H.~Leutwyler, A.~Pich and E.~de Rafael,
    Phys.\ Lett.\  B {\bf 223} (1989) 425.


\bibitem{NC}
  G.~'t Hooft,
  Nucl.\ Phys.\  B {\bf 72} (1974) 461;
%
    {\bf 75} (1974) 461;
%
  E.~Witten,
  Nucl.\ Phys.\  B {\bf 160} (1979) 57.


\bibitem{Pich}
    A. Pich,
    PoS Confinement  {\bf 8} (2008) 026     [arXiv:0812.2631 [hep-ph]]
    and references therein.


\bibitem{Karol} K.~Kampf and B.~Moussallam, Eur.\ Phys.\ J.\  C {\bf 47} (2006) 723


\bibitem{Estim}
    V.~Cirigliano {\it et al.},
    Nucl.\ Phys.\  B {\bf 753} (2006) 139



\bibitem{L8-Peris}
    M. Golterman and S. Peris,
    Phys. Rev. D {\bf 74} (2006) 096002.


\bibitem{MHA}
    M. Knecht and E. de Rafael,
    Phys. Lett. B {\bf 424} (1998) 335-342.


\bibitem{Pade}
    P. Masjuan and S. Peris,
    JHEP {\bf 0705} (2007) 040.


\bibitem{Weinberg2} S.~Weinberg, Phys.\ Rev.\ Lett.\  {\bf 18} (1967) 507.


\bibitem{prop} K.~Kampf, J.~Novotny and J.~Trnka, Fizika B {\bf 17} (2008) 349;
Nucl.\ Phys.\ Proc.\ Suppl.\  {\bf 186} (2009) 153;
arXiv:0905.1348 [hep-ph]; in preparation


\bibitem{QFT}
    O.~Cata and S.~Peris, Phys.\ Rev.\  D {\bf 65} (2002) 056014;
    I.~Rosell {\it et al.},
    JHEP {\bf 0408} (2004) 042.

\bibitem{form}
    I.~Rosell {\it et al.},
    JHEP {\bf 0701} (2007) 039;
    JHEP {\bf 0807} (2008) 014.


\bibitem{Group}
    J.~J.~Sanz-Cillero,
    Phys. Lett. B {\bf 681} (2009) 100-104.


\bibitem{Our}
    J.~J.~Sanz-Cillero and J.~Trnka,
    [arXiv:0912.0495 [hep-ph]].



\bibitem{NLOsatura}
    I. Rosell,
    P. Ruiz-Femen\'\i a and  J.J. Sanz-Cillero,
    Phys. Rev. D {\bf 79}  (2009) 076009.

\bibitem{Peris}
    M. Golterman and S. Peris,
    Phys. Rev. D {\bf 61} (2000) 034018.


\end{thebibliography}
\end{document}